\documentclass[fleqn,usenatbib]{mnras}

\usepackage{newtxtext,newtxmath}
\usepackage[T1]{fontenc}
\usepackage{ae,aecompl}
\usepackage{graphicx}	
\usepackage{amsmath}	
\usepackage{amssymb}

\title[The Relation Between Halo Mass and V$_{\rm flat}$]{The Tight Empirical Relation between Dark Matter Halo Mass and Flat Rotation Velocity for Late-Type Galaxies}

\author[H. Katz et al. ]{Harley Katz$^{1}$\thanks{E-mail: harley.katz@physics.ox.ac.uk}, Harry Desmond$^1$\thanks{St. John's JRF}, Stacy McGaugh$^2$ and Federico Lelli$^3$\thanks{ESO Fellow}\\
$^1$Astrophysics, University of Oxford, Denys Wilkinson Building, Keble Road, Oxford OX1 3RH, UK\\
$^2$Case Western Reserve University, Department of Astronomy, Cleveland OH, 44106, USA\\
$^3$European Southern Observatory, Karl-Schwarzschild-Strasse 2, Garching bei Munchen, Germany\\
}

\pubyear{2015}

\begin{document}
\label{firstpage}
\pagerange{\pageref{firstpage}--\pageref{lastpage}}
\maketitle

\begin{abstract}
We present a new empirical relation between galaxy dark matter halo mass (${\rm M_{halo}}$) and the velocity along the flat portion of the rotation curve (${\rm V_{flat}}$), derived from 120 late-type galaxies from the SPARC database.  The orthogonal scatter in this relation is comparable to the observed scatter in the baryonic Tully-Fisher relation (BTFR), indicating a tight coupling between total halo mass and galaxy kinematics at $r\ll R_{\rm vir}$.  The small vertical scatter in the relation makes it an extremely competitive estimator of total halo mass.  We demonstrate that this conclusion holds true for different priors on $M_*/L_{[3.6\mu]}$ that give a tight BTFR, but requires that the halo density profile follows DC14 rather than NFW.  We provide additional relations between ${\rm M_{halo}}$ and other velocity definitions at smaller galactic radii (i.e. ${\rm V_{2.2}}$, ${\rm V_{eff}}$, and ${\rm V_{max}}$) which can be useful for estimating halo masses from kinematic surveys, providing an alternative to abundance matching.  Furthermore, we constrain the dark matter analog of the Radial Acceleration Relation and also find its scatter to be small, demonstrating the fine balance between baryons and dark matter in their contribution to galaxy kinematics.
\end{abstract}


\section{Introduction}
The Tully--Fisher relation \citep[TFR;][]{Tully1977} was first formulated as a relation between optical luminosity and 21 cm line width in late-type galaxies as a way of accurately measuring distances to galaxies. Subsequently, it has been recognized that the line width is a proxy for the galaxy circular velocity \citep[e.g.][]{Verheijen2001} while the luminosity is a proxy for the stellar mass of the system. This led to the development of numerous alternative forms for the relation, replacing luminosity with stellar, gas, or total baryonic mass, and line width by the velocity at specific radii on a rotation curve (RC); such as the maximum rotation velocity ${\rm V_{max}}$, the velocity ${\rm V_{80}}$ at the radius enclosing 80\% of the light, or the velocity ${\rm V_{flat}}$ where the RC plateaus \citep[e.g.][]{McGaugh2000,Verheijen2001,ghasp,McGaugh2012}. These relations have proven useful not only for calibrating the cosmic distance ladder at low redshift (e.g.~\citealt{Tully_distances}), but also for providing a testing ground for models of galaxy formation (e.g.~\citealt{Eisenstein1996,Mo1998,Courteau1999,McGaugh2000,vdb2000,Mo2004,TG, Dutton10, Dutton11, Desmond_TFR}).

Since the luminosity of the galaxy is set by the baryonic matter and the rotation velocity is often dominated by the dark matter \citep[e.g.][]{Rubin1980,deBlok2001,deBlok2008}, the TFR provides a unique insight into the relation between these two components.  Interestingly, this relation extends more than six decades in baryonic mass while the intrinsic scatter is small and may be consistent with zero \citep{McGaugh2012,Lelli2016b}.  This is considered a strong test of $\Lambda$CDM due to the scatters expected between halo mass, concentration, and baryonic mass \citep{Maccio2008,Dutton2014,Desmond_BTFR}. Similarly, various dynamical processes can restructure the halo and gas distribution and impact ${\rm V_{flat}}$, such as adiabatic contraction \citep[e.g.][]{Blumenthal1986}, feedback-driven outflows \citep[e.g.][]{Navarro1996,Pontzen2012}, and dynamical friction \citep[e.g.][]{Elzant2001,Weinberg2002,Johansson2009}.  It is difficult to conceive of a scenario where the scatter in the BTFR remains consistent with zero when all of these processes nonlinearly interact and impact ${\rm V_{flat}}$.  Semi-analytic models that aim to understand the scatter in the BTFR are only marginally consistent with the observations \citep{Dutton2012, Desmond_BTFR}. Since ${\rm V_{flat}}$ is often dominated by halo mass, one might expect that the scatter in the ${\rm M_{halo}}-{\rm V_{flat}}$ relation would be less than that of the BTFR.  If this holds true, the kinematics of the galaxy at $r\ll R_{\rm vir}$ can be used to accurately estimate halo mass, providing an empirical alternative to other techniques such as abundance matching.  Until now, the ${\rm M_{halo}}-{\rm V_{flat}}$ relation has yet to be empirically determined, nor has the scatter been calculated. In this Letter we measure this relation and compare it with the BTFR for the same galaxy sample.

\begin{figure*}
\centerline{\includegraphics[scale=0.85,trim={0 0 0 0.5cm},clip]{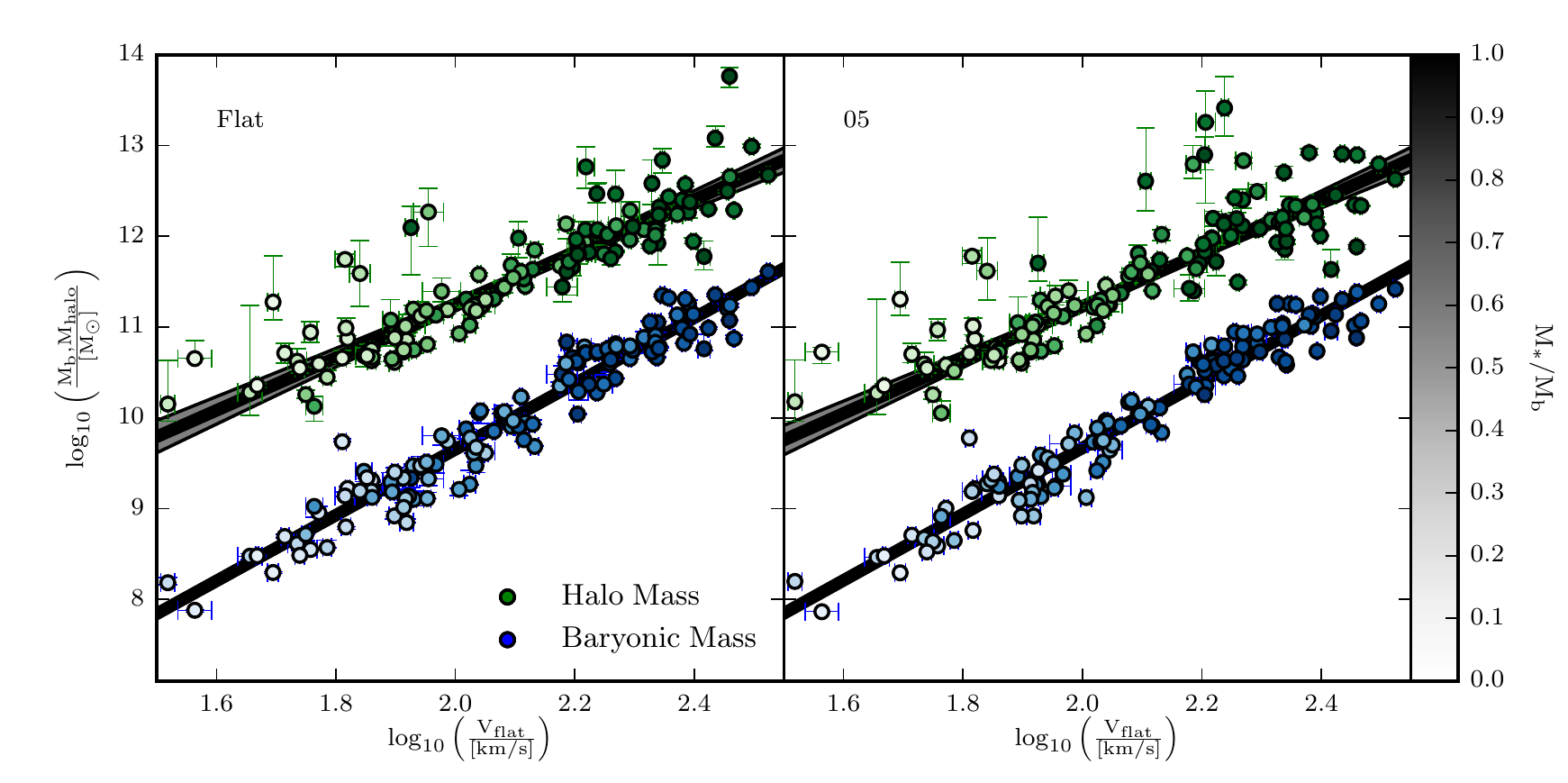}}
\caption{BTFR (blue) and ${\rm M_{halo}-V_{flat}}$ relation (green) for the Flat (left) or 05 (right) priors. Data points represent the maximum {\it a posteriori} values and the error bars show the $1\sigma$ highest posterior density confidence intervals on the 1D marginalized posterior. The thick black line and gray bands represent the best fit relations and $2\sigma$ confidence intervals calculated using the RANSAC algorithm. The points are shaded by their stellar-to-baryonic mass fractions.}
\label{TF}
\end{figure*}

\section{The ${\rm M_{halo}}-{\rm V_{flat}}$ and Baryonic Tully--Fisher relations}

To determine the ${\rm M_{halo}}$-${\rm V_{flat}}$ relation\footnote{We define ${\rm M_{halo}}$ to be the dark matter mass within the radius (${\rm R_{vir}}$) that contains an average density of 93.6$\rho_{\rm crit}$, consistent with a WMAP3 cosmology \citep{Spergel2007}.  ${\rm M_{vir}=M_{halo}+M_b}$.} and the BTFR, we employ the gas and stellar mass models from the SPARC data set \citep{Lelli2016} as well as the technique from  \cite{Katz2017}\footnote{The fitting procedure of \protect{\cite{Katz2017}} is also used to calculate the stellar mass-to-light ratio for each galaxy, given the priors, which determines the stellar mass.  Only the models without the ``$\Lambda$CDM" priors are used in the work.} to empirically determine halo masses.  We have rerun all of the Markov Chain Monte Carlo (MCMC) simulations presented in \cite{Katz2017} and now directly sample in $\log_{10}(\rm M_{halo})$, $\log_{10}(\rm c_{halo})$, and $\log_{10}(M_*/L)$, but all further analysis remains the same. Halo masses were estimated using both the DC14 halo profile \citep{DC2014}, a result from cosmological hydrodynamics simulations that can exhibit cusps or cores depending on the ratio of ${\rm M_*/M_{halo}}$, as well as the NFW halo profile \citep{NFW}, which is derived from dark matter-only simulations.  Only the former is consistent with other observational constraints \citep{Katz2017,Allaert2017}.  The NFW model often provides unrealistically large halo masses for galaxies with slowly rising rotation curves (the cusp-core problem) so that they fall very far from the stellar mass-halo mass relation derived from abundance matching (see Figure~3 of \citealt{Katz2017}). Of the 147 galaxies fit by \cite{Katz2017}, only 120 have measured values for ${\rm V_{flat}}$ (according to the \citealt{Lelli2016b} definition) and thus, only these galaxies are used to derive the ${\rm M_{halo}-V_{flat}}$ relation and the BTFR.

An important consideration is the choice of prior on the mass-to-light ratio ($M_*/L$), which impacts the inferred stellar, baryonic, and fitted halo masses. For our fiducial model we assume a flat prior in the range $0.3-0.8$ \citep[][designated as ``Flat'']{McGaugh2014}, although we also show results for the case where $M_*/L=0.5$ (``05''), which minimizes the scatter in the BTFR \citep{Lelli2016b}\footnote{We have also investigated the use of Gaussian Popsynth \citep{McGaugh2014} and DiskMass \citep{Martinsson2013} priors (see \citealt{Katz2017}); however these result in BTFRs that have much larger scatter compared to the 05 prior so we do not consider them in the remainder of our analysis.  However, see Table~\ref{TFtable} for the fitted values and scatters of the BTFR and ${\rm M_{halo}}$-${\rm V_{flat}}$ relation.}.

In Fig~\ref{TF}, we show the resulting ${\rm M_{halo}-V_{flat}}$ relation for both $M_*/L$ priors and the corresponding BTFRs. Note how the higher mass galaxies have a higher ratio of stellar mass to total baryonic mass (see also \citealt{Katz2018}).  In order to constrain the mean relations, we must account for the non-Gaussian and asymmetric uncertainties on ${\rm M_{halo}}$ and the observational error bars on V$_{\rm flat}$. To do this, we create 10,000 re-sampled catalogs where we randomly draw a halo mass for each galaxy from the posterior distribution mapped out by the MCMC chains and a ${\rm V_{flat}}$ from a Gaussian distribution using the measured ${\rm V_{flat}}$ from the RC and its uncertainty.

Because of uncertainties on galaxy distance, systematic features in the RCs\footnote{These may result from non-circular motion or asymmetries.}, and the limited radial extent of the RCs\footnote{This can lead to large uncertainties in halo mass if the halo RC is still rising out to the last measured point.}, we expect that not all galaxies will be well fit by the halo model and therefore outliers will be present in our dataset. As these may bias our fits to the relations, we employ the RANSAC algorithm \citep{Fischler1981}. This is a robust estimation technique that uses an iterative procedure to determine whether a data point is an inlier or an outlier given the other data in the set, without relying on sigma clipping. We use RANSAC to fit the ${\rm M_{halo}-V_{flat}}$ relation and the BTFR for each of the 10,000 catalogs using the following equation:
\begin{equation}
    \log_{10}({\rm M_{halo\ or\ b}/M_\odot})=A\log_{10}({\rm V_{flat}}/\text{km}\:\text{s}^{-1})+B.
    \label{HMVR}
\end{equation}
The mean and standard deviations of the fit parameters are calculated using the 10,000 realizations and the resulting relations and $1\sigma$ confidence intervals are shown as the black line and gray shaded regions in Fig.~\ref{TF}.

\begin{table*}
\centering
\begin{tabular}{@{}lllllllllllll@{}}
$M_*/L$ & Halo & Mass & Velocity & $A$ & $\sigma_A$ & $B$ & $\sigma_B$ & $\sigma_{\perp}$ & $\sigma_{\perp}$ &  $\sigma_{\rm M_b,M_{halo}}$ & $\sigma_{\rm M_b,M_{halo}}$ & Outlier\\
Prior & Model & Measure & Measure & & & & & (inliers) & (all) & (inliers) & (all) &  Fraction\\
 \hline

Flat & DC14 & ${\rm M_ b}$ & ${\rm V_{flat}}$ & 3.623 & 0.023 & 2.406 & 0.051 & 0.062 & 0.062 & 0.234 & 0.234 & 0.001 \\
Flat & DC14 & ${\rm M_{halo}}$ & ${\rm V_{flat}}$ & 2.902 & 0.138 & 5.439 & 0.292 & 0.064 & 0.075 & 0.195 & 0.231 & 0.124 \\

05 & DC14 & ${\rm M_{b}}$ & ${\rm V_{flat}}$ & 3.647 & 0.024 & 2.374 & 0.055 & 0.062 & 0.063 & 0.236 & 0.237 & 0.004 \\
05 & DC14 & ${\rm M_{\rm halo}}$ & ${\rm V_{flat}}$ & 2.947 & 0.136 & 5.334 & 0.283 & 0.067 & 0.082 & 0.207 & 0.256 & 0.139 \\

Flat & NFW & ${\rm M_{b}}$ & ${\rm V_{flat}}$ & 3.630 & 0.028 & 2.400 & 0.062 & 0.060 & 0.060 & 0.225 & 0.226 & 0.003 \\
Flat & NFW & ${\rm M_{halo}}$ & ${\rm V_{flat}}$ & 2.216 & 0.208 & 6.907 & 0.471 & 0.116 & 0.193 & 0.279 & 0.466 & 0.347 \\

Flat & DC14 & ${\rm M_{b}}$ & ${\rm V_{2.2}}$ & 2.805 & 0.064 & 4.216 & 0.139 & 0.089 & 0.090 & 0.264 & 0.269 & 0.016 \\
Flat & DC14 & ${\rm M_{halo}}$ & ${\rm V_{2.2}}$ & 2.083 & 0.211 & 7.218 & 0.450 & 0.121 & 0.147 & 0.277 & 0.336 & 0.151 \\

05 & DC14 & ${\rm M_{b}}$ & ${\rm V_{2.2}}$ & 2.859 & 0.104 & 4.108 & 0.211 & 0.080 & 0.084 & 0.241 & 0.254 & 0.042 \\
05 & DC14 & ${\rm M_{halo}}$ & ${\rm V_{2.2}}$ & 2.276 & 0.162 & 6.800 & 0.340 & 0.107 & 0.134 & 0.266 & 0.332 & 0.146 \\

Flat & DC14 & ${\rm M_{b}}$ & ${\rm V_{eff}}$ & 2.474 & 0.051 & 5.016 & 0.110 & 0.102 & 0.104 & 0.271 & 0.277 & 0.020 \\
Flat & DC14 & ${\rm M_{halo}}$ & ${\rm V_{eff}}$ & 1.877 & 0.166 & 7.707 & 0.353 & 0.130 & 0.167 & 0.275 & 0.353 & 0.172 \\

05 & DC14 & ${\rm M_{b}}$ & ${\rm V_{eff}}$ & 2.468 & 0.051 & 5.054 & 0.109 & 0.100 & 0.101 & 0.266 & 0.270 & 0.013 \\
05 & DC14 & ${\rm M_{halo}}$ & ${\rm V_{eff}}$ & 1.868 & 0.190 & 7.742 & 0.397 & 0.134 & 0.174 & 0.283 & 0.367 & 0.173 \\

Flat & DC14 & ${\rm M_{b}}$ & ${\rm V_{max}}$ & 3.204 & 0.080 & 3.214 & 0.177 & 0.072 & 0.074 & 0.242 & 0.249 & 0.024 \\
Flat & DC14 & ${\rm M_{halo}}$ & ${\rm V_{max}}$ & 2.439 & 0.196 & 6.333 & 0.424 & 0.093 & 0.113 & 0.244 & 0.297 & 0.139 \\

05 & DC14 & ${\rm M_{b}}$ & ${\rm V_{max}}$ & 3.253 & 0.096 & 3.123 & 0.211 & 0.072 & 0.074 & 0.246 & 0.252 & 0.019 \\
05 & DC14 & ${\rm M_{halo}}$ & ${\rm V_{max}}$ & 2.348 & 0.277 & 6.538 & 0.595 & 0.106 & 0.131 & 0.267 & 0.328 & 0.150 \\
\hline
DiskMass & DC14 & ${\rm M_{b}}$ & ${\rm V_{flat}}$ & 3.259 & 0.063 & 2.986 & 0.127 & 0.086 & 0.089 & 0.293 & 0.302 & 0.032 \\
DiskMass & DC14 & ${\rm M_{halo}}$ & ${\rm V_{flat}}$ & 2.596 & 0.067 & 5.982 & 0.147 & 0.062 & 0.067 & 0.173 & 0.187 & 0.052 \\

Popsynth & DC14 & ${\rm M_{b}}$ & ${\rm V_{flat}}$ & 3.509 & 0.157 & 2.583 & 0.307 & 0.072 & 0.076 & 0.261 & 0.277 & 0.054 \\
Popsynth & DC14 & ${\rm M_{halo}}$ & ${\rm V_{flat}}$ & 2.773 & 0.071 & 5.652 & 0.153 & 0.062 & 0.076 & 0.182 & 0.224 & 0.128 \\

\hline
\end{tabular}
\caption{Best-fit parameters and their $1\sigma$ uncertainties for the BTFR and ${\rm M_{halo}-V}$ relations as given in Equation ~\ref{HMVR}. For the DC14 profile we show results for both the Flat and 05 priors (as well as DiskMass and Popsynth for ${\rm V_{flat}}$), and for the NFW profile only the former.  $\sigma$ denotes scatter orthogonal to the best-fit line, derived by multiplying the mean of the MADs of the 10,000 fits by 1.48.  We present the scatters both with (all) and without (inliers) outliers for both the orthogonal and vertical scatters to the relation.}
\label{TFtable}
\end{table*}

The best-fitting parameters for the ${\rm M_{halo}-V_{flat}}$ relation and the BTFR, and their uncertainties, are listed in Table~\ref{TFtable} for both sets of priors.  We also quote the scatter in the relations in the $M$-direction and the orthogonal ($\perp$) scatter quantified by 1.48 times the median absolute deviations (MADs) of the points from the best-fitting lines \citep{McGaugh2012}, and the percentage of outliers identified by RANSAC\footnote{We have checked that most outliers are a result of probabilistically sampling the wide posterior in halo mass because the halo RC is still rising out to the furthest observed radius.}. Interestingly, we find that the $\perp$ scatter in the ${\rm M_{halo}-V_{flat}}$ relation is comparable to that of the BTFR for both $M_*/L$ priors {\rm within $1\sigma$} (but see Section~\ref{cavs}).  The vertical scatter is then smaller in the ${\rm M_{halo}-V_{flat}}$ relation compared to the BTFR due to the shallower slope. The $\perp$ scatter may indicate which relation is more fundamental while the vertical scatter can be used to determine the accuracy of the mass prediction from ${\rm V_{flat}}$.  For comparison, we also present in Table~\ref{TFtable} the results when using the NFW halo profile: it is evident that both the scatter in the ${\rm M_{halo}-V_{flat}}$ relation and the outlier fractions are significantly increased.  This demonstrates that having a tight ${\rm M_{halo}-V_{flat}}$ relation is not guaranteed when fitting the RCs with any halo model.

Since the ${\rm M_{halo}-V_{flat}}$ relation has more outliers than the BTFR, we have also computed the $\perp$ and vertical scatters using all of the galaxies in each catalog (i.e. assuming no outliers), which puts an upper limit on the scatters in the relations. Even in this extreme case, which almost certainly overestimates the scatter, the $\perp$ scatters in the BTFR and ${\rm M_{halo}-V_{flat}}$ relation are comparable to within $2\sigma$, and the vertical scatters are nearly identical, regardless of the prior on $M_*/L$.  However, this is not the case for NFW. Our estimated slopes and normalizations for the BTFR are unsurprisingly independent of halo profile and are very consistent with the error-weighted fits from \cite{Lelli2016b} who use the same data set with a fixed mass-to-light ratio.  Likewise, the slopes are slightly shallower, albeit still consistent within the uncertainties with the estimates from \cite{McGaugh2012}.  The observed scatter we measure in the BTFR is also entirely consistent with \cite{McGaugh2012} and \cite{Lelli2016b}.

\begin{figure*}
\centerline{\includegraphics[scale=1,trim={0 0.1cm 0 0.2cm},clip]{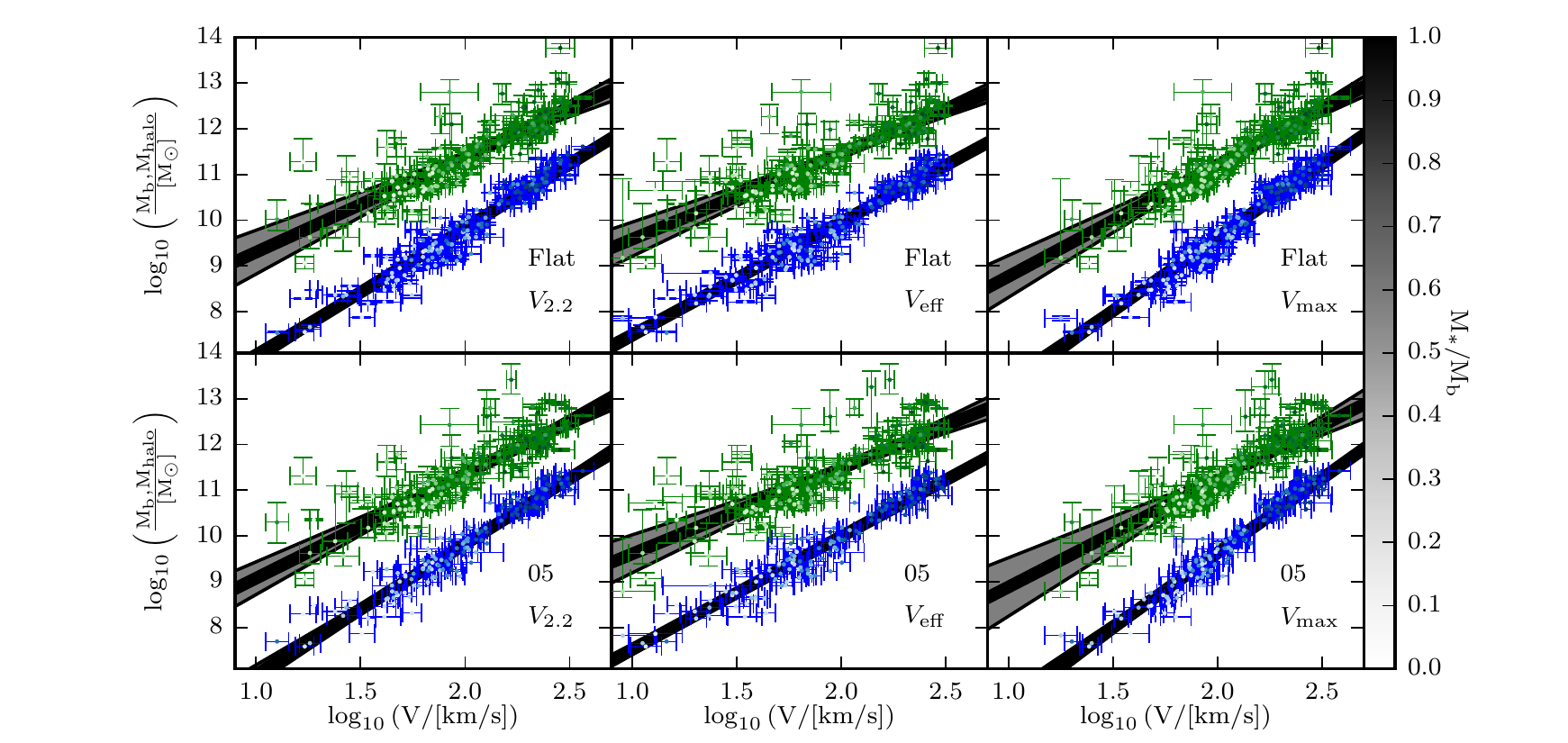}}
\caption{BTFR (blue) and ${\rm M_{halo}-V}$ relation (green) for different velocity measures (indicated on panels). Lines, error bars and shading are as in Fig.~\ref{TF}. The fit parameters are listed in Table~\ref{TFtable}.}
\label{TF_All}
\end{figure*}

\begin{figure}
\centerline{\includegraphics[scale=1,trim={0 0.7cm 0 0.2cm},clip]{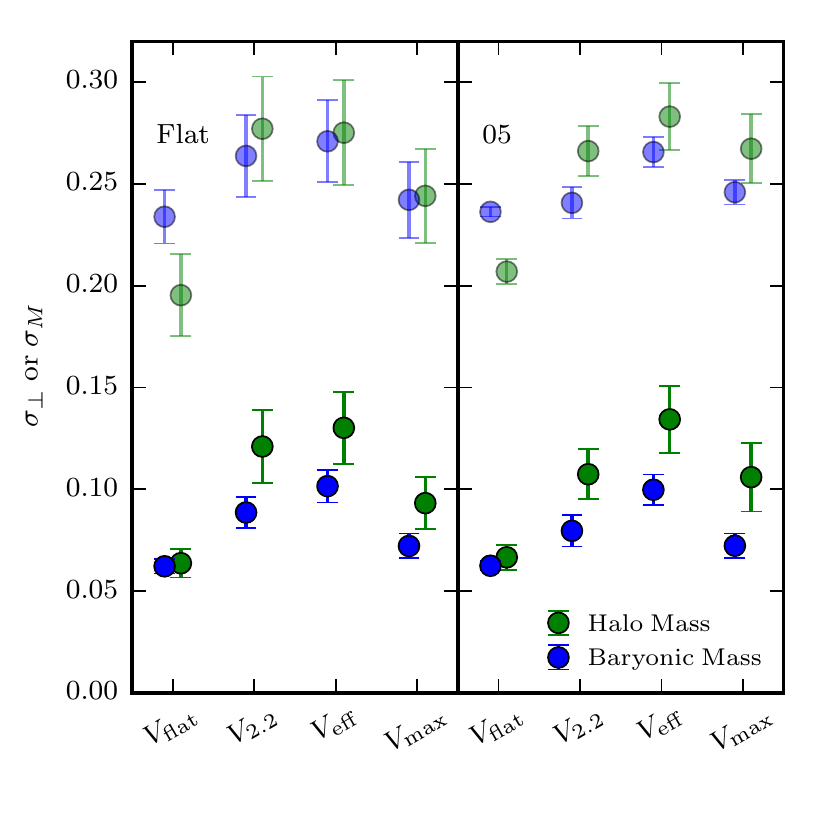}}
\caption{The scatter in the $\perp$ (dark) and $M$ (light) directions of the BTFR (blue) or ${\rm M_{halo}-V}$ relations (green) for different velocity measures. The points show the average over the 10,000 model realisations, and the error bars show 1.48 times the MAD. The left and right panels are for the Flat and 05 priors, respectively.  See Section~\ref{cavs} for caveats in directly comparing the scatters.}
\label{scatters}
\end{figure}

\begin{figure}
\centerline{\includegraphics[scale=1,trim={0 0.5cm 0 0.2cm},clip]{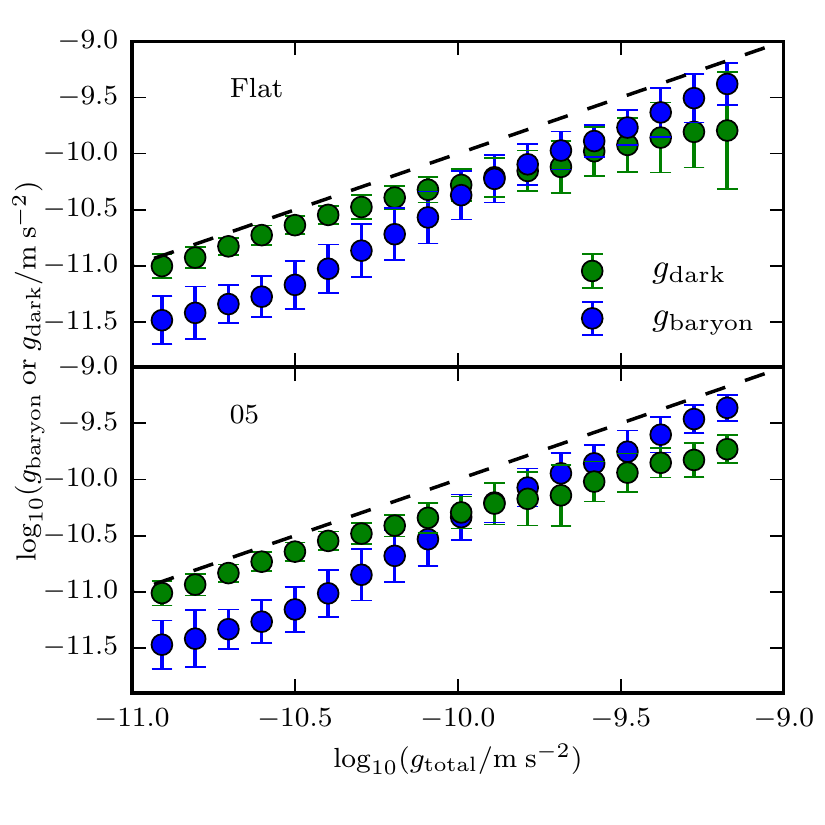}}
\caption{Relation between $g_\text{total}$ and $g_\text{baryon}$ (Radial Acceleration Relation; blue) and between $g_\text{total}$ and $g_\text{dark}$ (halo Radial Acceleration Relation; green) for the Flat (upper) and 05 (bottom) priors. The dashed line indicates $x=y$.}
\label{rarscatters}
\end{figure}

The derived ${\rm M_{halo}-V_{flat}}$ relation is considerably shallower than the BTFR, with a slope slightly less than 3 indicating that the ${\rm V_{halo}-V_{flat}}$ relation has a slope near unity, where ${\rm V_{halo}}$ is the circular velocity of the dark matter halo at ${\rm R_{vir}}$.  Directly measuring this relation, we find:
\begin{subequations}
\begin{align}
{\rm Flat:\  V_{halo}=(0.904\pm0.216)V_{flat}^{0.947\pm0.051}}\\
{\rm 05:\ V_{halo}=(1.028\pm0.251)V_{flat}^{0.919\pm0.052}}
\end{align}
\end{subequations} 
with $\sigma_{\rm V_{vir}}=0.06,0.07$~dex for the Flat and 05 priors, respectively, and hence ${\rm V_{halo}\sim V_{flat}}$.

We also consider several common choices for the velocity measure used to define the relations.  Although our fiducial choice is ${\rm V_{flat}}$, we also investigate ${\rm V_{2.2}}$ (measured at 2.2 disk scale lengths), ${\rm V_{max}}$, and ${\rm V_{eff}}$ (at the half-light radius).  The disk scale length and the half-light radius are measured for each galaxy in the SPARC data set from the surface brightness profile (see Section~3.1 of \citealt{Lelli2016}).  We show the BTFR and ${\rm M_{halo}-V}$ relations for all of these velocity measures in Fig.~\ref{TF_All}, and quantify their scatters in Fig.~\ref{scatters}. The error bars are the standard deviations across the 10,000 Monte Carlo realizations. In nearly all cases, the scatters in the BTFR and ${\rm M_{halo}-V}$ relations are consistent within $\sim1-1.5\sigma$ (but see Section~\ref{cavs}). However for these other velocity measures, the $\perp$ scatter for the  ${\rm M_{halo}-V}$ relations tend to fall systematically above the BTFRs.  This is not surprising: ${\rm V_{flat}}$ is generally measured furthest out in the galaxy where the halo often dominates the RC while the others are measured much closer to the center of the galaxy. However, ${\rm V_{flat}}$ also gives the tightest BTFR and is therefore likely to be the most fundamental velocity to use in this context. In Table~\ref{TFtable} we list the best-fit values of the fit parameters in each case, including both Flat and 05 priors on $M_*/L$, as well as the outlier fraction and scatter $\perp$ to the best-fit line. These relations allow the halo mass to be determined (up to an uncertainty given by $\sigma$) for any galaxy in which any of these velocities has been measured.

\section{The halo Radial Acceleration Relation}
\label{hrar}

Recently,~\citet{OneLaw, RAR} reported a tight correlation between the centripetal acceleration due to baryons, $g_\text{baryon}$, and the total acceleration $g_\text{total}$, across the RCs of the SPARC galaxies (the Radial Acceleration Relation, RAR). This relation is a re-parametrization of the mass discrepancy--acceleration relation~\citep{Sanders, MG_1999, MG_MDAR}. The RAR shows that at high $g_\text{baryon}$, $g_\text{total}$ and $g_\text{baryon}$ are approximately equal, so that baryons are the only dynamically-relevant mass component. Conversely, at low $g_\text{baryon}$, $g_\text{total}/g_\text{baryon}$ rises, indicating the increasing prominence of dark matter. The tightness of the relation is surprising given the expected scatter in the galaxy--halo connection \citep{Desmond_MDAR}, and indicates that galaxy RCs may be well estimated from only the baryon distribution.

Our halo fits to the SPARC RCs allow us to construct a related relation, the \emph{halo Radial Acceleration Relation} (hRAR). Instead of $g_\text{baryon}$, this plots $g_\text{dark}$ against $g_\text{total}$: we derive $g_\text{dark}$ by sampling the RC of each SPARC galaxy at the same radii as~\citet{Lelli2016} and calculating $g_\text{dark} = V_{\rm halo}^2(r)/r$. As above, we take 10,000 Monte Carlo realizations of the uncertainties for either the Flat or 05 prior, and bin the result in $g_\text{total}$. The results are compared with the RAR in Fig.~\ref{rarscatters}, removing bins with fewer than 50 points per model realization. The error bars show the $1\sigma$ standard deviation in $g_\text{baryon}$ or $g_\text{dark}$, and the dashed black line shows $g_\text{dark}$\ {\rm or}\ $g_\text{baryon}$=$g_\text{total}$.

The trends are qualitatively as expected: at lower $g_\text{total}$, dark matter accounts for almost all of the acceleration, so that the uncertainty on $g_\text{dark}$ at fixed $g_\text{total}$ is small. In contrast, for the Flat $M_*/L$ prior, at high acceleration $g_\text{total}$ is a poor predictor of enclosed dark matter mass because baryons dominate. The relations cross over at $g_\text{total} \sim 10^{-10}$ m/s$^2$, indicating the transition between baryon and dark matter dominance. We see however that the choice of $M_*/L$ prior has a large effect on the scatter of the relations, especially the hRAR at high $g_\text{total}$. For $M_*/L = 0.5$, $g_\text{dark}$ may in fact be estimated precisely from $g_\text{total}$ even in high-acceleration regions; almost as precisely as $g_\text{baryon}$. Although $g_\text{dark}$ must be inferred from the RCs themselves -- and hence the hRAR does not relate observables as the RAR does -- it does fill in the part of galaxy dynamics that the RAR misses, and demonstrates more explicitly the relative role the two mass components play in setting the kinematics of different parts of galaxies.

Finally, we quantify the total scatter in the hRAR relative to the RAR \citep{Desmond_MDAR}.  For the Flat prior we find $\sigma_\text{tot, RAR} = 0.208$ and $\sigma_\text{tot, hRAR} = 0.168$, while for the 05 prior we find $\sigma_\text{tot, RAR} = 0.201$ and $\sigma_\text{tot, hRAR} = 0.138$. 

\section{Caveats}
\label{cavs}
It should be emphasized that although we have directly compared the BTFR with the ${\rm M_{halo}}-{\rm V_{flat}}$ relation for the same set of galaxies, the two variables of the BTFR have been determined by completely independent observations while both ${\rm V_{flat}}$ and ${\rm M_{halo}}$ are measured from the same RCs.  It is very difficult to determine how much the estimate of ${\rm M_{halo}}$ is dependent on ${\rm V_{flat}}$.  We find that the dark matter contribution to the square of the galaxy circular velocity at ${\rm R_{flat}}$, the innermost radius where the RC becomes flat (see Section~2.2 of \citealt{Lelli2016b} for details on the derivation of ${\rm V_{flat}}$), is $56\%\pm24\%$ (1$\sigma$ standard deviation) with a weak correlation for higher mass galaxies having a smaller contribution.  Furthermore, the fraction of halo mass that exists at $r<{\rm R_{flat}}$ can range between $<1\%$ to 35\% with a mean of $\sim7\%$, indicating that the majority of the halo mass is at much larger radii than where ${\rm V_{flat}}$ is measured.  The outer slope of the DC14 profile, which helps set the virial mass, is also dependent on the stellar content of the galaxy.  While there is indeed a covariance between the two quantities, which may reduce the scatter of the ${\rm M_{halo}}-{\rm V_{flat}}$ relation, we stress again that this exercise does not work for the NFW halo (as shown in Table~\ref{TFtable}) and is unlikely to work for any arbitrary density profile (i.e. pseudo-isothermal, logarithmic, etc.).  The DC14 model not only provides good fits to the RCs while simultaneously being in agreement with estimates of the ${\rm M_*-M_{halo}}$ relation from abundance matching and mass-concentration relation from dark matter-only simulations \citep{Katz2017}, but now also produces a ${\rm M_{halo}}-{\rm V_{flat}}$ relation with scatter comparable to the BTFR.  The goal of our experiment is not to determine conclusively which relation is more fundamental, but rather to constrain the parameters and scatter of what appears to be a new and tight relation.

\section{Conclusions}
We present the ${\rm M_{halo}-V_{flat}}$ relation, empirically determined from the SPARC data set of late-type galaxies \citep{Lelli2016} and the corresponding dark matter halo fits from \cite{Katz2017}. This has a similar form to the baryonic Tully--Fisher relation (BTFR) and uses M$_\text{halo}$ rather than M$_\text{b}$, but is more fundamental in the context of $\Lambda$CDM where galaxy dynamics in the outer regions are typically set by the halo mass.  We find that the orthogonal scatters are comparable between the two relations.  The ${\rm M_{halo}-V_\text{flat}}$ relation is well modeled by a power law over many decades in mass with a slope only slightly shallower than 3, indicating that ${\rm V_{halo}\sim V_{flat}}$. The small vertical scatter in the ${\rm M_{halo}-V_{flat}}$ relation means that halo masses can be accurately determined from measurements along the flat part of galaxy RCs. We repeat this exercise for multiple velocity measures to provide a means of estimating halo mass from a variety of kinematical observations.  We also present the halo analogue of the Radial Acceleration Relation, replacing baryonic by dark matter acceleration, and show that it has small scatter and a well-defined shape. These relations characterize the relative importance and balance of baryons and dark matter across the RCs of late-type galaxies.

\section*{Acknowledgements}
We thank the referee for their comments which improved the manuscript.  H.K. thanks the Beecroft fellowship, the Nicholas Kurti Junior Fellowship, and Brasenose College. H.D. is supported by St John's College, Oxford.

\bibliographystyle{mnras}
\bibliography{main.bib}

\begin{thebibliography}{}
\makeatletter
\relax
\def\mn@urlcharsother{\let\do\@makeother \do\$\do\&\do\#\do\^\do\_\do\%\do\~}
\def\mn@doi{\begingroup\mn@urlcharsother \@ifnextchar [ {\mn@doi@}
  {\mn@doi@[]}}
\def\mn@doi@[#1]#2{\def\@tempa{#1}\ifx\@tempa\@empty \href
  {http://dx.doi.org/#2} {doi:#2}\else \href {http://dx.doi.org/#2} {#1}\fi
  \endgroup}
\def\mn@eprint#1#2{\mn@eprint@#1:#2::\@nil}
\def\mn@eprint@arXiv#1{\href {http://arxiv.org/abs/#1} {{\tt arXiv:#1}}}
\def\mn@eprint@dblp#1{\href {http://dblp.uni-trier.de/rec/bibtex/#1.xml}
  {dblp:#1}}
\def\mn@eprint@#1:#2:#3:#4\@nil{\def\@tempa {#1}\def\@tempb {#2}\def\@tempc
  {#3}\ifx \@tempc \@empty \let \@tempc \@tempb \let \@tempb \@tempa \fi \ifx
  \@tempb \@empty \def\@tempb {arXiv}\fi \@ifundefined
  {mn@eprint@\@tempb}{\@tempb:\@tempc}{\expandafter \expandafter \csname
  mn@eprint@\@tempb\endcsname \expandafter{\@tempc}}}

\bibitem[\protect\citeauthoryear{{Allaert}, {Gentile}  \& {Baes}}{{Allaert}
  et~al.}{2017}]{Allaert2017}
{Allaert} F.,  {Gentile} G.,   {Baes} M.,  2017, \mn@doi [\aap]
  {10.1051/0004-6361/201730402}, \href
  {http://adsabs.harvard.edu/abs/2017A%26A...605A..55A} {605, A55}

\bibitem[\protect\citeauthoryear{{Blumenthal}, {Faber}, {Flores}  \&
  {Primack}}{{Blumenthal} et~al.}{1986}]{Blumenthal1986}
{Blumenthal} G.~R.,  {Faber} S.~M.,  {Flores} R.,   {Primack} J.~R.,  1986,
  \mn@doi [\apj] {10.1086/163867}, \href
  {http://adsabs.harvard.edu/abs/1986ApJ...301...27B} {301, 27}

\bibitem[\protect\citeauthoryear{{Courteau} \& {Rix}}{{Courteau} \&
  {Rix}}{1999}]{Courteau1999}
{Courteau} S.,  {Rix} H.-W.,  1999, \mn@doi [\apj] {10.1086/306872}, \href
  {http://adsabs.harvard.edu/abs/1999ApJ...513..561C} {513, 561}

\bibitem[\protect\citeauthoryear{{Desmond}}{{Desmond}}{2017a}]{Desmond_MDAR}
{Desmond} H.,  2017a, \mn@doi [\mnras] {10.1093/mnras/stw2571}, \href
  {http://adsabs.harvard.edu/abs/2017MNRAS.464.4160D} {464, 4160}

\bibitem[\protect\citeauthoryear{{Desmond}}{{Desmond}}{2017b}]{Desmond_BTFR}
{Desmond} H.,  2017b, \mn@doi [\mnras] {10.1093/mnrasl/slx134}, \href
  {http://adsabs.harvard.edu/abs/2017MNRAS.472L..35D} {472, L35}

\bibitem[\protect\citeauthoryear{{Desmond} \& {Wechsler}}{{Desmond} \&
  {Wechsler}}{2015}]{Desmond_TFR}
{Desmond} H.,  {Wechsler} R.~H.,  2015, \mn@doi [\mnras]
  {10.1093/mnras/stv1978}, \href
  {http://adsabs.harvard.edu/abs/2015MNRAS.454..322D} {454, 322}

\bibitem[\protect\citeauthoryear{{Di Cintio}, {Brook}, {Dutton}, {Macci{\`o}},
  {Stinson}  \& {Knebe}}{{Di Cintio} et~al.}{2014}]{DC2014}
{Di Cintio} A.,  {Brook} C.~B.,  {Dutton} A.~A.,  {Macci{\`o}} A.~V.,
  {Stinson} G.~S.,   {Knebe} A.,  2014, \mn@doi [\mnras]
  {10.1093/mnras/stu729}, \href
  {http://adsabs.harvard.edu/abs/2014MNRAS.441.2986D} {441, 2986}

\bibitem[\protect\citeauthoryear{{Dutton}}{{Dutton}}{2012}]{Dutton2012}
{Dutton} A.~A.,  2012, \mn@doi [\mnras] {10.1111/j.1365-2966.2012.21469.x},
  \href {http://adsabs.harvard.edu/abs/2012MNRAS.424.3123D} {424, 3123}

\bibitem[\protect\citeauthoryear{{Dutton} \& {Macci{\`o}}}{{Dutton} \&
  {Macci{\`o}}}{2014}]{Dutton2014}
{Dutton} A.~A.,  {Macci{\`o}} A.~V.,  2014, \mn@doi [\mnras]
  {10.1093/mnras/stu742}, \href
  {http://adsabs.harvard.edu/abs/2014MNRAS.441.3359D} {441, 3359}

\bibitem[\protect\citeauthoryear{{Dutton}, {Conroy}, {van den Bosch}, {Prada}
  \& {More}}{{Dutton} et~al.}{2010}]{Dutton10}
{Dutton} A.~A.,  {Conroy} C.,  {van den Bosch} F.~C.,  {Prada} F.,   {More} S.,
   2010, \mn@doi [\mnras] {10.1111/j.1365-2966.2010.16911.x}, \href
  {http://adsabs.harvard.edu/abs/2010MNRAS.407....2D} {407, 2}

\bibitem[\protect\citeauthoryear{{Dutton} et~al.,}{{Dutton}
  et~al.}{2011}]{Dutton11}
{Dutton} A.~A.,  et~al., 2011, \mn@doi [\mnras]
  {10.1111/j.1365-2966.2011.19038.x}, \href
  {http://adsabs.harvard.edu/abs/2011MNRAS.416..322D} {416, 322}

\bibitem[\protect\citeauthoryear{{Eisenstein} \& {Loeb}}{{Eisenstein} \&
  {Loeb}}{1996}]{Eisenstein1996}
{Eisenstein} D.~J.,  {Loeb} A.,  1996, \mn@doi [\apj] {10.1086/176905}, \href
  {http://adsabs.harvard.edu/abs/1996ApJ...459..432E} {459, 432}

\bibitem[\protect\citeauthoryear{{El-Zant}, {Shlosman}  \& {Hoffman}}{{El-Zant}
  et~al.}{2001}]{Elzant2001}
{El-Zant} A.,  {Shlosman} I.,   {Hoffman} Y.,  2001, \mn@doi [\apj]
  {10.1086/322516}, \href {http://adsabs.harvard.edu/abs/2001ApJ...560..636E}
  {560, 636}

\bibitem[\protect\citeauthoryear{Fischler \& Bolles}{Fischler \&
  Bolles}{1981}]{Fischler1981}
Fischler M.~A.,  Bolles R.~C.,  1981, \mn@doi [Commun. ACM]
  {10.1145/358669.358692}, 24, 381

\bibitem[\protect\citeauthoryear{{Johansson}, {Naab}  \&
  {Ostriker}}{{Johansson} et~al.}{2009}]{Johansson2009}
{Johansson} P.~H.,  {Naab} T.,   {Ostriker} J.~P.,  2009, \mn@doi [\apjl]
  {10.1088/0004-637X/697/1/L38}, \href
  {http://adsabs.harvard.edu/abs/2009ApJ...697L..38J} {697, L38}

\bibitem[\protect\citeauthoryear{{Katz}, {Lelli}, {McGaugh}, {Di Cintio},
  {Brook}  \& {Schombert}}{{Katz} et~al.}{2017}]{Katz2017}
{Katz} H.,  {Lelli} F.,  {McGaugh} S.~S.,  {Di Cintio} A.,  {Brook} C.~B.,
  {Schombert} J.~M.,  2017, \mn@doi [\mnras] {10.1093/mnras/stw3101}, \href
  {http://adsabs.harvard.edu/abs/2017MNRAS.466.1648K} {466, 1648}

\bibitem[\protect\citeauthoryear{{Katz}, {Desmond}, {Lelli}, {McGaugh}, {Di
  Cintio}, {Brook}  \& {Schombert}}{{Katz} et~al.}{2018}]{Katz2018}
{Katz} H.,  {Desmond} H.,  {Lelli} F.,  {McGaugh} S.,  {Di Cintio} A.,  {Brook}
  C.,   {Schombert} J.,  2018, \mn@doi [\mnras] {10.1093/mnras/sty2129}, \href
  {http://adsabs.harvard.edu/abs/2018MNRAS.480.4287K} {480, 4287}

\bibitem[\protect\citeauthoryear{{Lelli}, {McGaugh}  \& {Schombert}}{{Lelli}
  et~al.}{2016a}]{Lelli2016}
{Lelli} F.,  {McGaugh} S.~S.,   {Schombert} J.~M.,  2016a, \mn@doi [\aj]
  {10.3847/0004-6256/152/6/157}, \href
  {http://adsabs.harvard.edu/abs/2016AJ....152..157L} {152, 157}

\bibitem[\protect\citeauthoryear{{Lelli}, {McGaugh}  \& {Schombert}}{{Lelli}
  et~al.}{2016b}]{Lelli2016b}
{Lelli} F.,  {McGaugh} S.~S.,   {Schombert} J.~M.,  2016b, \mn@doi [\apjl]
  {10.3847/2041-8205/816/1/L14}, \href
  {http://adsabs.harvard.edu/abs/2016ApJ...816L..14L} {816, L14}

\bibitem[\protect\citeauthoryear{{Lelli}, {McGaugh}, {Schombert}  \&
  {Pawlowski}}{{Lelli} et~al.}{2017}]{OneLaw}
{Lelli} F.,  {McGaugh} S.~S.,  {Schombert} J.~M.,   {Pawlowski} M.~S.,  2017,
  \mn@doi [\apj] {10.3847/1538-4357/836/2/152}, \href
  {http://adsabs.harvard.edu/abs/2017ApJ...836..152L} {836, 152}

\bibitem[\protect\citeauthoryear{{Macci{\`o}}, {Dutton}  \& {van den
  Bosch}}{{Macci{\`o}} et~al.}{2008}]{Maccio2008}
{Macci{\`o}} A.~V.,  {Dutton} A.~A.,   {van den Bosch} F.~C.,  2008, \mn@doi
  [\mnras] {10.1111/j.1365-2966.2008.14029.x}, \href
  {http://adsabs.harvard.edu/abs/2008MNRAS.391.1940M} {391, 1940}

\bibitem[\protect\citeauthoryear{{Martinsson}, {Verheijen}, {Westfall},
  {Bershady}, {Andersen}  \& {Swaters}}{{Martinsson}
  et~al.}{2013}]{Martinsson2013}
{Martinsson} T.~P.~K.,  {Verheijen} M.~A.~W.,  {Westfall} K.~B.,  {Bershady}
  M.~A.,  {Andersen} D.~R.,   {Swaters} R.~A.,  2013, \mn@doi [\aap]
  {10.1051/0004-6361/201321390}, \href
  {http://adsabs.harvard.edu/abs/2013A%26A...557A.131M} {557, A131}

\bibitem[\protect\citeauthoryear{{McGaugh}}{{McGaugh}}{1999}]{MG_1999}
{McGaugh} S.,  1999, in {Merritt} D.~R.,  {Valluri} M.,   {Sellwood} J.~A.,
  eds,  Astronomical Society of the Pacific Conference Series Vol. 182, Galaxy
  Dynamics - A Rutgers Symposium.

\bibitem[\protect\citeauthoryear{{McGaugh}}{{McGaugh}}{2004}]{MG_MDAR}
{McGaugh} S.~S.,  2004, \mn@doi [\apj] {10.1086/421338}, \href
  {http://adsabs.harvard.edu/abs/2004ApJ...609..652M} {609, 652}

\bibitem[\protect\citeauthoryear{{McGaugh}}{{McGaugh}}{2012}]{McGaugh2012}
{McGaugh} S.~S.,  2012, \mn@doi [\aj] {10.1088/0004-6256/143/2/40}, \href
  {http://adsabs.harvard.edu/abs/2012AJ....143...40M} {143, 40}

\bibitem[\protect\citeauthoryear{{McGaugh} \& {Schombert}}{{McGaugh} \&
  {Schombert}}{2014}]{McGaugh2014}
{McGaugh} S.~S.,  {Schombert} J.~M.,  2014, \mn@doi [\aj]
  {10.1088/0004-6256/148/5/77}, \href
  {http://adsabs.harvard.edu/abs/2014AJ....148...77M} {148, 77}

\bibitem[\protect\citeauthoryear{{McGaugh}, {Schombert}, {Bothun}  \& {de
  Blok}}{{McGaugh} et~al.}{2000}]{McGaugh2000}
{McGaugh} S.~S.,  {Schombert} J.~M.,  {Bothun} G.~D.,   {de Blok} W.~J.~G.,
  2000, \mn@doi [\apjl] {10.1086/312628}, \href
  {http://adsabs.harvard.edu/abs/2000ApJ...533L..99M} {533, L99}

\bibitem[\protect\citeauthoryear{{McGaugh}, {Lelli}  \& {Schombert}}{{McGaugh}
  et~al.}{2016}]{RAR}
{McGaugh} S.~S.,  {Lelli} F.,   {Schombert} J.~M.,  2016, \mn@doi [Physical
  Review Letters] {10.1103/PhysRevLett.117.201101}, \href
  {http://adsabs.harvard.edu/abs/2016PhRvL.117t1101M} {117, 201101}

\bibitem[\protect\citeauthoryear{{Mo} \& {Mao}}{{Mo} \& {Mao}}{2004}]{Mo2004}
{Mo} H.~J.,  {Mao} S.,  2004, \mn@doi [\mnras]
  {10.1111/j.1365-2966.2004.08114.x}, \href
  {http://adsabs.harvard.edu/abs/2004MNRAS.353..829M} {353, 829}

\bibitem[\protect\citeauthoryear{{Mo}, {Mao}  \& {White}}{{Mo}
  et~al.}{1998}]{Mo1998}
{Mo} H.~J.,  {Mao} S.,   {White} S.~D.~M.,  1998, \mn@doi [\mnras]
  {10.1046/j.1365-8711.1998.01227.x}, \href
  {http://adsabs.harvard.edu/abs/1998MNRAS.295..319M} {295, 319}

\bibitem[\protect\citeauthoryear{{Navarro}, {Eke}  \& {Frenk}}{{Navarro}
  et~al.}{1996a}]{Navarro1996}
{Navarro} J.~F.,  {Eke} V.~R.,   {Frenk} C.~S.,  1996a, \mn@doi [\mnras]
  {10.1093/mnras/283.3.L72}, \href
  {http://adsabs.harvard.edu/abs/1996MNRAS.283L..72N} {283, L72}

\bibitem[\protect\citeauthoryear{{Navarro}, {Frenk}  \& {White}}{{Navarro}
  et~al.}{1996b}]{NFW}
{Navarro} J.~F.,  {Frenk} C.~S.,   {White} S.~D.~M.,  1996b, \mn@doi [\apj]
  {10.1086/177173}, \href {http://adsabs.harvard.edu/abs/1996ApJ...462..563N}
  {462, 563}

\bibitem[\protect\citeauthoryear{{Pontzen} \& {Governato}}{{Pontzen} \&
  {Governato}}{2012}]{Pontzen2012}
{Pontzen} A.,  {Governato} F.,  2012, \mn@doi [\mnras]
  {10.1111/j.1365-2966.2012.20571.x}, \href
  {http://adsabs.harvard.edu/abs/2012MNRAS.421.3464P} {421, 3464}

\bibitem[\protect\citeauthoryear{{Rubin}, {Ford}  \& {Thonnard}}{{Rubin}
  et~al.}{1980}]{Rubin1980}
{Rubin} V.~C.,  {Ford} Jr. W.~K.,   {Thonnard} N.,  1980, \mn@doi [\apj]
  {10.1086/158003}, \href {http://adsabs.harvard.edu/abs/1980ApJ...238..471R}
  {238, 471}

\bibitem[\protect\citeauthoryear{{Sanders}}{{Sanders}}{1990}]{Sanders}
{Sanders} R.~H.,  1990, \mn@doi [\aapr] {10.1007/BF00873540}, \href
  {http://adsabs.harvard.edu/abs/1990A%26ARv...2....1S} {2, 1}

\bibitem[\protect\citeauthoryear{{Spergel} et~al.,}{{Spergel}
  et~al.}{2007}]{Spergel2007}
{Spergel} D.~N.,  et~al., 2007, \mn@doi [\apjs] {10.1086/513700}, \href
  {http://adsabs.harvard.edu/abs/2007ApJS..170..377S} {170, 377}

\bibitem[\protect\citeauthoryear{{Torres-Flores}, {Epinat}, {Amram}, {Plana}
  \& {Mendes de Oliveira}}{{Torres-Flores} et~al.}{2011}]{ghasp}
{Torres-Flores} S.,  {Epinat} B.,  {Amram} P.,  {Plana} H.,   {Mendes de
  Oliveira} C.,  2011, \mn@doi [\mnras] {10.1111/j.1365-2966.2011.19169.x},
  \href {http://adsabs.harvard.edu/abs/2011MNRAS.416.1936T} {416, 1936}

\bibitem[\protect\citeauthoryear{{Trujillo-Gomez}, {Klypin}, {Primack}  \&
  {Romanowsky}}{{Trujillo-Gomez} et~al.}{2011}]{TG}
{Trujillo-Gomez} S.,  {Klypin} A.,  {Primack} J.,   {Romanowsky} A.~J.,  2011,
  \mn@doi [\apj] {10.1088/0004-637X/742/1/16}, \href
  {http://adsabs.harvard.edu/abs/2011ApJ...742...16T} {742, 16}

\bibitem[\protect\citeauthoryear{{Tully} \& {Fisher}}{{Tully} \&
  {Fisher}}{1977}]{Tully1977}
{Tully} R.~B.,  {Fisher} J.~R.,  1977, \aap, \href
  {http://adsabs.harvard.edu/abs/1977A%26A....54..661T} {54, 661}

\bibitem[\protect\citeauthoryear{{Tully} \& {Pierce}}{{Tully} \&
  {Pierce}}{2000}]{Tully_distances}
{Tully} R.~B.,  {Pierce} M.~J.,  2000, \mn@doi [\apj] {10.1086/308700}, \href
  {http://adsabs.harvard.edu/abs/2000ApJ...533..744T} {533, 744}

\bibitem[\protect\citeauthoryear{{Verheijen}}{{Verheijen}}{2001}]{Verheijen2001}
{Verheijen} M.~A.~W.,  2001, \mn@doi [\apj] {10.1086/323887}, \href
  {http://adsabs.harvard.edu/abs/2001ApJ...563..694V} {563, 694}

\bibitem[\protect\citeauthoryear{{Weinberg} \& {Katz}}{{Weinberg} \&
  {Katz}}{2002}]{Weinberg2002}
{Weinberg} M.~D.,  {Katz} N.,  2002, \mn@doi [\apj] {10.1086/343847}, \href
  {http://adsabs.harvard.edu/abs/2002ApJ...580..627W} {580, 627}

\bibitem[\protect\citeauthoryear{{de Blok}, {McGaugh}  \& {Rubin}}{{de Blok}
  et~al.}{2001}]{deBlok2001}
{de Blok} W.~J.~G.,  {McGaugh} S.~S.,   {Rubin} V.~C.,  2001, \mn@doi [\aj]
  {10.1086/323450}, \href {http://adsabs.harvard.edu/abs/2001AJ....122.2396D}
  {122, 2396}

\bibitem[\protect\citeauthoryear{{de Blok}, {Walter}, {Brinks}, {Trachternach},
  {Oh}  \& {Kennicutt}}{{de Blok} et~al.}{2008}]{deBlok2008}
{de Blok} W.~J.~G.,  {Walter} F.,  {Brinks} E.,  {Trachternach} C.,  {Oh}
  S.-H.,   {Kennicutt} Jr. R.~C.,  2008, \mn@doi [\aj]
  {10.1088/0004-6256/136/6/2648}, \href
  {http://adsabs.harvard.edu/abs/2008AJ....136.2648D} {136, 2648}

\bibitem[\protect\citeauthoryear{{van den Bosch}}{{van den
  Bosch}}{2000}]{vdb2000}
{van den Bosch} F.~C.,  2000, \mn@doi [\apj] {10.1086/308337}, \href
  {http://adsabs.harvard.edu/abs/2000ApJ...530..177V} {530, 177}

\makeatother
\end{thebibliography}

\label{lastpage}
\end{document}